\documentclass[10pt,twocolumn]{article}

\usepackage{amsmath}
\usepackage{amssymb}
\usepackage{isotope}
\usepackage{lipsum} 		
\usepackage{blindtext} 		

\usepackage[superscript, biblabel]{cite}  	

\makeatletter
\renewcommand\@biblabel[1]{#1.}
\makeatother

\usepackage{etoolbox}
\patchcmd{\thebibliography}{\section*{\refname}}{}{}{}

\usepackage[margin=1.0in,hmarginratio=1:1,top=32mm,columnsep=15pt]{geometry}

\usepackage{color}				
\definecolor{dark-gray}{gray}{0.1}	
\usepackage{times}				
\usepackage{microtype} 			
\usepackage[english]{babel} 		

\usepackage[hang, font={footnotesize}, labelfont={bf,up}, textfont={sf,up}]{caption} 
\usepackage{booktabs} 	
\usepackage{mwe}  		
\usepackage{graphicx}  		

\usepackage{enumitem} 		
\setlist[itemize]{noitemsep} 	

\usepackage{abstract} 
\AtBeginDocument{}  		

\usepackage{titlesec} 			
\renewcommand\thesection{\Roman{section}.} 		  		
\renewcommand\thesubsection{\thesection\Alph{subsection}.} 	
\renewcommand\thesubsubsection{\thesubsection\arabic{subsubsection}.} 
\titleformat{\section}[block]{\normalfont\sffamily\bfseries}{\thesection}{1em}{\MakeUppercase}{} 	
\titleformat{\subsection}[block]{\normalfont\sffamily\bfseries}{\thesubsection}{1em}{}{}  
\titleformat{\subsubsection}[block]{\normalfont\sffamily\bfseries}{\thesubsubsection}{1em}{}{}  
\titlespacing*{\section}{0.0em}{1em}{0.25em}		
\titlespacing*{\subsection}{0.0em}{1em}{0.25em}	
\usepackage{indentfirst}						

\usepackage{tocloft}
\usepackage{fancyhdr} 

\fancypagestyle{plain}{
\fancyhf{}
\lhead{\color{dark-gray}\textit{}} 
\rhead{} 
}

\usepackage{titling} 		

\usepackage{hyperref} 	
\hypersetup{
	backref=true,       
    	pagebackref=true,               
    	hyperindex=true,                
    	colorlinks=true,                
    	breaklinks=true,                
    	urlcolor= blue,                
     	linkcolor=blue,  
    	linkcolor=blue,                
    	bookmarks=true,                 
    	bookmarksopen=false,
    	filecolor=black,
    	citecolor=blue
}

\pretitle{\begin{center}\large\bfseries} 	
\posttitle{\end{center}} 				
\fancypagestyle{plain}{
\fancyhf{}
\lhead{\color{dark-gray}\textit{}} 
\rhead{ {\it  Los Alamos National Laboratory report LA-UR-22-32324 (2022)  } }
}

\title{\vspace{-0.3in} \sffamily{The earliest  DT nuclear fusion discoveries}}
\author{%
\normalsize M.\ B.\ Chadwick\thanks{corresponding author: mbchadwick@lanl.gov}, G.\ M.\ Hale,  M.\ W.\ Paris, J.\ P.\ Lestone, C. Bates, 
J.\ B.\ Wilhelmy, and S.\ A.\  Andrews \\[-0.5ex]
\normalsize Los Alamos National Laboratory, Los Alamos, NM 87545 \\
\normalsize W.\ Tornow and  S.\ W.\ Finch \\[-0.5ex] 
\normalsize Triangle Universities Nuclear Laboratory and Duke University, Durham, NC 27708\\
}
\date{ } 
\renewcommand{\maketitlehookd}{%
\begin{abstract}
\vspace*{-0.5in}
\noindent {\normalfont\textbf{Abstract}\textemdash}
\noindent 

We describe the earliest measurements of the DT fusion cross section commissioned by the Manhattan Project, first at Purdue University in 1943 and then at Los Alamos 1945-6 and later, in 1951-2. The Los Alamos measurements led to the realization that a 3/2$^+$ resonance in the DT system enhances the fusion cross section {\it by a factor of one hundred} at energies relevant to applications. This was a transformational discovery, making the quest for terrestrial fusion energy possible.
The earliest measurements were reasonably accurate given the technology of the time and the scarcity of tritium, and were quickly improved
   to provide cross section data accurate to just a few percent. We provide a previously-unappreciated insight: that DT fusion was first reported in Ruhlig's 1938 University of Michigan experiment and likely influenced Konopinski in 1942 to suggest its usefulness for thermonuclear technologies.    
   We report on preliminary work to repeat the 1938 measurement, and our simulations of that experiment.
   We also present some work by Fermi, from his 1945 Los Alamos lectures, showing that he used the S-factor concept about a decade before it was introduced by 
nuclear astrophysicists.

\end{abstract}
}

\begin{document}

\maketitle	


\section{Introduction} 

Fusion energy research has seen exciting recent breakthroughs. Livermore's NIF laser facility has achieved ignition\cite{NIF:2022,Clery:2022b} and Culham's Joint European Torus (JET) has produced a record 59 MJ of fusion energy\cite{Clery:2022}. Against this backdrop of advances, we provide an account of the earliest fusion discoveries from the 1930s --1950s. Some of this technical history has not been previously appreciated, most notably the first 1938 reporting of DT 14 MeV neutrons at the University of Michigan by Arthur Ruhlig\cite{Ruhlig:1938}. We argue here that this experiment had a critical role in inspiring early thermonuclear fusion research directions. Our paper presents some unique insights from the extensive holdings within Los Alamos National Laboratory's archives – including sources typically unavailable to a broad audience. This paper provides a short summary of a more comprehensive article under development\cite{Chadwick:2022}.
 
The first breakthroughs in fusion came almost two decades before the discovery of fission. In 1920 Eddington\cite{Eddington:1920} in
Cambridge proposed that stellar energy came from the fusion of hydrogen into helium, following Aston's measurements of the mass differences. 
Later that decade, Gamow published his $\alpha$-decay paper\cite{gamow:1928qta} that applied quantum mechanical tunneling to understand
nuclear decay rates. Gamow then showed that the same concepts could be adapted for understanding the inverse process of alpha particle
fusion\cite{gamow:1929zqa}. This led Houtermans, together with Atkinson, a British scientist working in G\"{o}ttingen who remembered
Eddington’s estimates of star temperatures, to suggest that proton nuclear fusion reactions powered stars\cite{atkinson:1929tle}.

\begin{figure}[htbp]
\begin{center}
\includegraphics[width=2.6in]{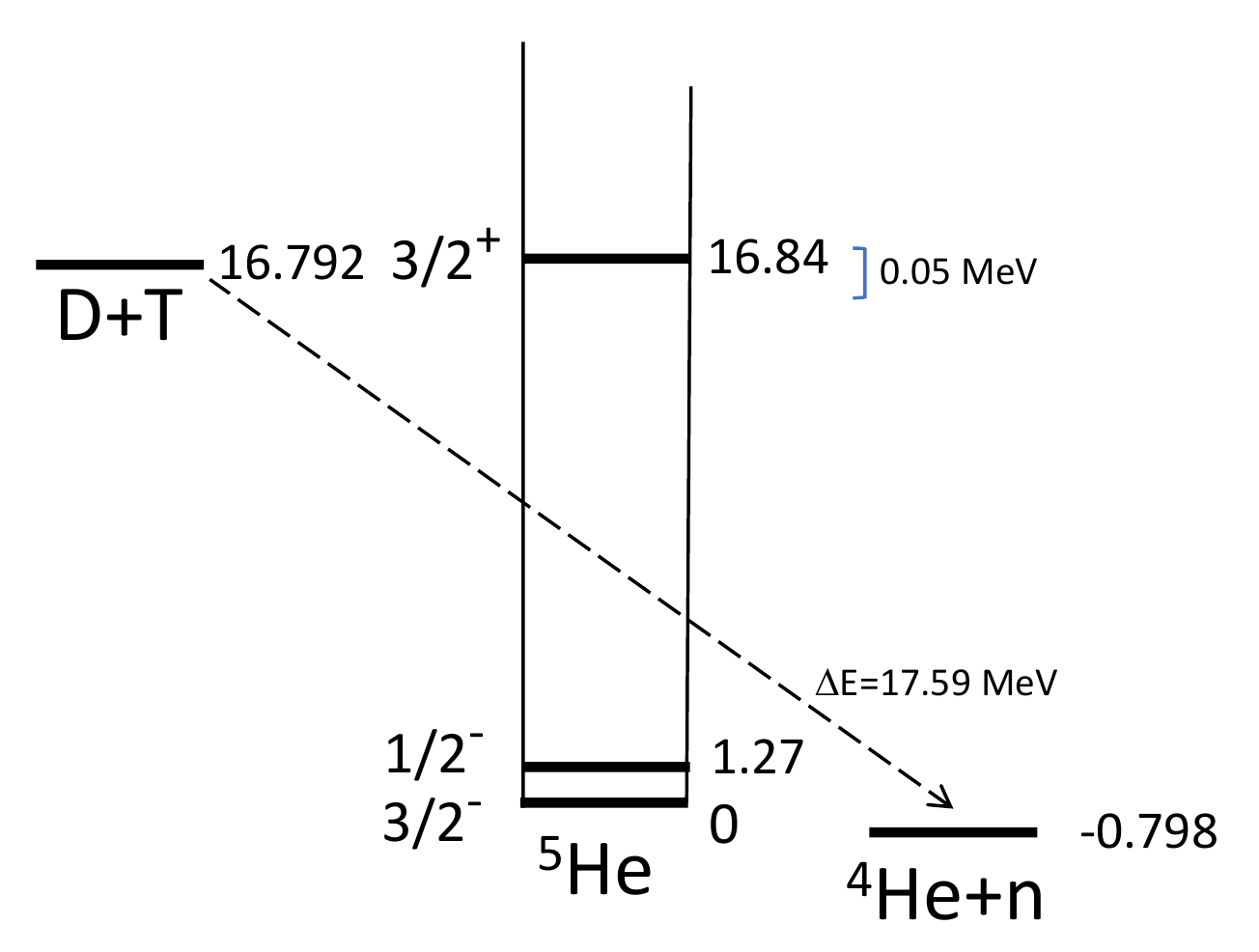}
\caption{Energy levels (in MeV) in the DT system making a compound system A=5. The threshold for DT reactions is seen to be very similar to the 3/2$^+$ resonance energy.}
\label{fig:A=5levels}
\end{center}
\vspace{-1.cm}
\end{figure}
\vspace{0.5cm}

\begin{figure*}[htbp]
\begin{center}
   \includegraphics[width=\textwidth]{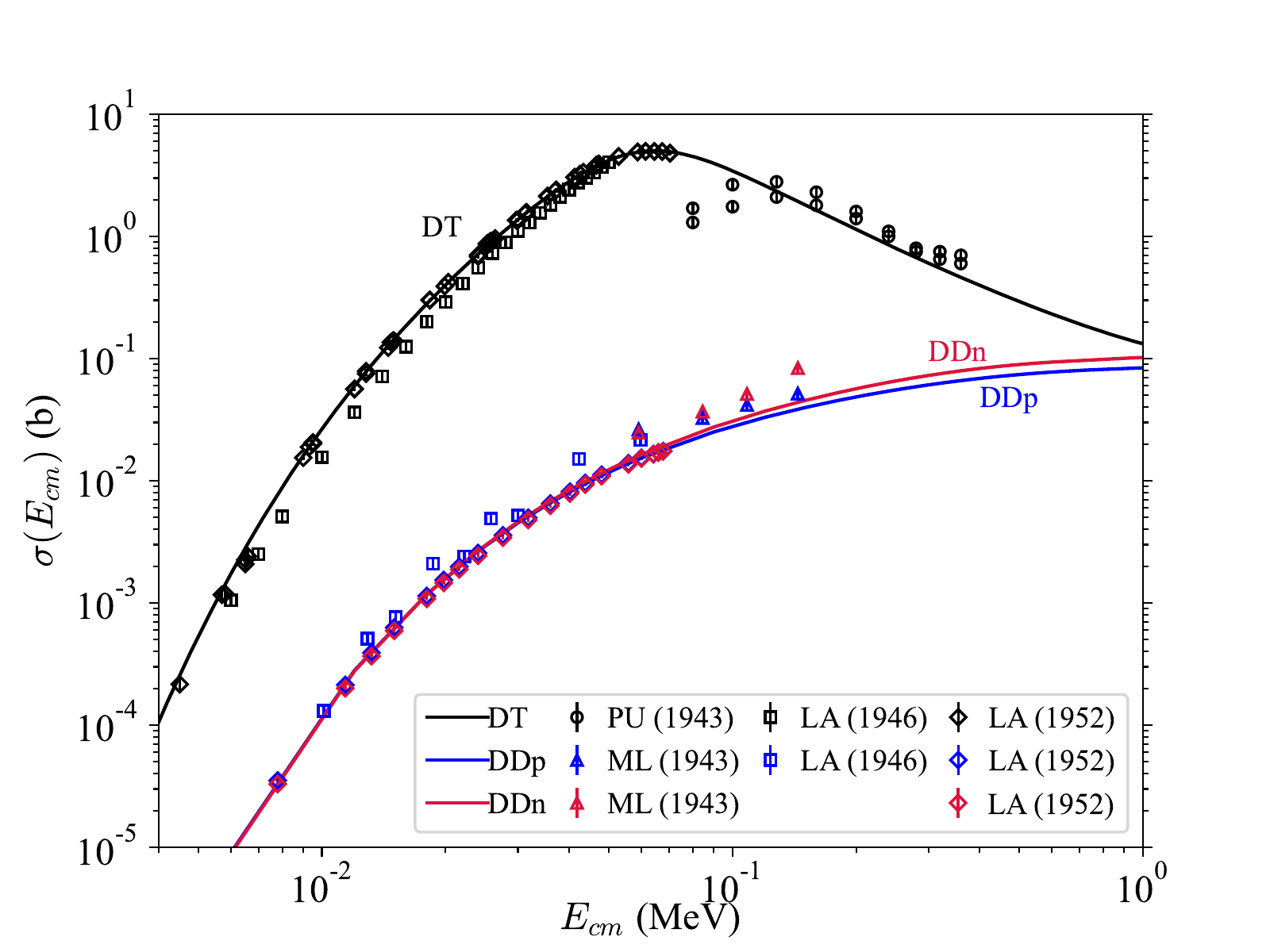} \caption{DT (black curve), DDn (red curve) and DDp (blue curve) fusion
   cross sections compared with 1940s-1950s data, and with the ENDF/B-VIII.0 evaluation (solid curves) from the $R$-matrix analysis of all
   known data. The DT data are from Purdue University (PU, 1943; two sets -- see text)\cite{baker:1943tr}, Los Alamos
   (LA,1946)\cite{Bretscher:1946a,Bretscher:1949} and Los Alamos (LA,1952)\cite{apsst:1952dtn,apsst:1954pr}.  The DDn data are  the Chicago
   Met Lab (ML, 1943)\cite{Coon:1944n,Bethe:1943b} and Los Alamos (LA,1952)\cite{apsst:1954pr}.  The DDp data are  the Chicago Met Lab (ML,
   1943)\cite{Coon:1944p}, Los Alamos (LA,1946)\cite{Bretscher:1946b} and Los Alamos (LA,1952)\cite{apsst:1954pr}.  The $R$-matrix analysis
   is, additionally, strongly constrained by other data (not shown), including the very accurate ($\lesssim 1.5\%$) 1980s - 1990 data by Jarmie, Brown
   {\it et al.}\cite{jarmie:1984prc,Brown:1990}.}
   \label{fig:dt-dd-cf-efb80}
\end{center}
\vspace{-1.cm}
\end{figure*}

The 1943-1946  Manhattan Project observations -- that the DT fusion cross section exceeds the DD cross sections by a factor of one hundred, described below --  came as a great surprise. It was a transformational discovery, opening up the possibility of terrestrial fusion energy production, and today  hot plasma DT reactions remain the primary focus  for accomplishing controlled fusion energy. Our account here features the discovery of the important role of the 3/2$^+$ resonance in the DT reaction (Fig.~\ref{fig:A=5levels}), causing this fusion cross section to be so high, Fig.~\ref{fig:dt-dd-cf-efb80}. Until measurements were obtained, it would not have been possible to have predicted that the 3/2$^+$ excited state in the A=5 system, at 16.84 MeV, would be at just the right energy needed to largely enhance the T(d,n)$\alpha$ fusion rate. The separation energy, 16.792 MeV, for the DT system is very close, providing a strong resonant enhancement for DT plasma temperatures around 10 keV (as observed in the NIF N210808 experiment\cite{NIF:2022}). It is one of those great coincidences in nature (like the location of the $^{12}$C 7.65 MeV state that enhances 3$\alpha$ nucleosynthesis\cite{Chadwick:2015}), and one that has 
substantial implications for the future peaceful production of controlled fusion energy.

\section{The first  DD and DT observations in the 1930s} 

The DD fusion reactions were first observed in 1934 by 
Oliphant, Harteck and Rutherford\cite{Oliphant:1934} at Cambridge's Cavendish laboratory, using the new 
Cockcroft and Walton accelerator.  They measured both the D(d,n)$^3$He and D(d,p)T reactions, which we refer to here  simply as DDn and DDp. They 
correctly found that the two branches are about equal, and noted the large amounts of fusion energy released, about 3 and 4 MeV respectively for each branch. Having discovered tritium and measured the DD reaction Q-values, the paper could have inferred the very large Q value for the important T(d,n)$\alpha$ fusion reaction. It didn't, although we have shown\cite{Chadwick:2022} that information given 
in that paper would lead one to calculate $\sim$18.3 MeV, close to the 17.6 MeV we know today, of which 14 MeV goes to the neutron and the remaining energy to the alpha particle. 

Although the DD reactions were widely studied at laboratories in Europe and America in the 1930s, an understanding of fusion reactions involving tritium proved elusive. Tritium was simply unavailable. Just before Rutherford died, he wrote a paper\cite{Rutherford:1937} wondering why tritium and $^3$He are so hard to find in 
nature, speculating that their cross sections are large and they disappear through nuclear reactions. While a prescient insight, it was not founded on compelling data.
 It was not until 1939 that Alvarez made enough tritium, via the DDp reaction in the Berkeley cyclotron, to determine that it is radioactive. After Fermi achieved criticality with the first reactor 
 at Chicago in December, 1942, a new production path for tritium via $^6$Li(n,t) opened up, 
 but this took two years to put into practice because reactor production of plutonium was a
 higher priority. 
 
 Our work has uncovered a strangely-neglected 1938 Physical Review paper: Ruhlig's\cite{Ruhlig:1938} letter to the editor, which contains --
 we think -- the first-ever reporting of DT fusion neutrons. The main focus of Ruhlig's short paper was studies of the DDn reaction
 using 0.5 MeV deuterons on a heavy phosphoric acid (D$_3$PO$_4$) target, using the University of Michigan's high-voltage ion tube
 accelerator. In brief comments
 at the end, this same paper notes that he also observed 15+ MeV neutrons. Ruhlig speculated that these came from {\it secondary} reactions in which, first,
 D(d,p)T reactions occur, after which some of the tritons (born at an energy of about 1 MeV) undergo secondary T(d,n)$\alpha$ nuclear
 reactions as they slow down. 
 
 This was a remarkably clever insight! 1n 1938, tritium could not be made in sufficient quantity  for direct DT fusion cross section
 studies, and yet Ruhlig was able to infer a large  D(T,n)$\alpha$ fusion rate. We are unaware of any citations to Ruhlig's seminal
 observation in the published literature: according to Google Scholar, until now it has only been cited seven times between 1939 and 1945
 but never for its DT observation. 
(In recent years NIF\cite{Cerjan:2018,Sayre:2019} has used such secondary DT neutron production in 
 DD ICF capsules to study plasma stopping powers, see the Appendix.)
 
 As we shall discuss further below, though, there is strong circumstantial evidence that this paper was
 known to scientists who came to Los Alamos in 1943 for the Manhattan Project, including Konopinski and Bethe, and influenced
 early optimism regarding the feasibility of developing an H-bomb. Perhaps it was not cited in the early days owing to wartime secrecy
 concerns, and then it subsequently remained unappreciated and became lost in time.
 
 Ruhlig's paper provides a numerical value for  his observation of secondary DT neutron production that we can check, given our modern understanding. He measured the 
  n-p proton recoils observed in his cloud chamber, finding a ratio of 1:1000 for the high energy DT recoils to
  those from the lower energy DD reactions. Based 
 on this measurement he  suggested that the DT reaction is ``exceedingly probable''. Today we can calculate the triton slowing down process with DT reactions through our knowledge of stopping powers and cross sections. Lestone {\it et al.} obtain\cite{Lestone:2022} a ratio more like10$^{-6}$--10$^{-5}$ (depending on assumptions made) instead of 10$^{-3}$. It is not clear to us why Ruhlig reported a higher number;  Lestone {\it et al.} have undertaken a detailed analysis of the experiment and present some possibilities\footnote{Lestone calculates the number of DT reactions for a triton born at 1.1 MeV as 4.8x10$^{-5}$ (confirmed by a separate FLAG simulation; also we thank George Zimmerman for an independent Livermore calculation that gave a similar result, 3.6x10$^{-5}$). To compare to Ruhlig's recoil proton ratio,  this is 
 reduced by the relative  efficiencies of n-p recoil detection for DTn v. DDn, together with the requirement that the DTn n-p proton recoils exceed the 15 MeV threshold to penetrate the carbon foil.  With varying plausible assumptions of what exactly was observed for the lower  and the higher energy recoils, the calculated ratio can 
 vary substantially.  But Lestone {\it et al.}\cite{Lestone:2022} also think it is possible that most of the high energy protons instead came -- not from neutron-induced  proton recoils -- but from the parallel  process D(D,$^3$He)n. That is, 
 secondary  $^{3}$He(D,p)$\alpha$ reactions  make high energy $\sim$15 MeV protons. Independent of these speculations, 
Ruhlig did produce {\it some} DT neutrons: the Michigan ion tube could deliver 250 $\mu$A, although on this target it presumably ran at a lower current; if he ran at 10 $\mu$A  (as Oliphant did, earlier) we calculate he would have observed tens of high energy recoils from DT per hour.} but a firm conclusion is hampered by a lack of detail in Ruhlig's short paper. 
See the Appendix for a modern measurement of this process.

It seems to us that  Ruhlig  did observe DT fusion neutrons, as he claimed: when he placed a 4 cm piece of lead between the target and the detector (shielding all protons) he still saw $>$ 15 MeV proton recoil events. These most likely did come from DT neutrons above 15 MeV that created n-p recoils.

 \section{Konopinski's DT Proposal in Berkeley, 1942 } 
 
Edward Teller's decadal quest to develop a thermonuclear bomb began after Fermi suggested that a fission explosion could ignite fusion in deuterium,  at Columbia University in 1941. The next major step to advance fusion energy came at Berkeley in July 1942. Oppenheimer convened a small meeting of  leading theoretical physicists, the ``luminaries'', to assess the feasibility of developing an atom bomb. 
The gathering included Bethe, Teller, Konopinski, Serber, van Vleck, 
Bloch, 
Frankel, 
and Nelson.  Tradition has it that the question at hand was answered positively and  disposed of quickly, and they spent the rest of the meeting thinking about the hydrogen bomb\cite{Bethe:1997}. As nuclear astrophysicists, the complex science involving thermonuclear reactions, hot plasmas and radiation transport, was a natural fascination for them. At that meeting, Konopinski suggested that the DT fusion process could be beneficial compared to DD.

 Teller wrote about this momentous insight in his 1955 Science magazine article  {\it The Work of Many People}
\cite{Teller:1955}: ``I remember particularly the suggestion of Konopinski that the reactions of tritium should be investigated. At that
time it was a mere guess. But it turned out to be an inspired one.''  However, we have found that Teller's characterization as a ``guess" is not quite right. 

Los Alamos'  archive contains an audio tape of an interview with Konopinski from 1986 \cite{Konopinski:1986}. When asked to describe his
suggestion to use tritium, Konopinski replied\cite{Chadwick:2022}:
``That was way back in Berkeley... I happened to know from pre-war work that the reaction of deuterium with tritium produces much
more energy and has a larger cross section than D+D. ...''. In the course of our studies, at first we thought Konopinski must have mis-remembered this, because the DT cross section was not measured until {\it later}, in 1943 at Purdue University. But then we discovered the aforementioned 1938 paper by
Ruhlig, where a large DT reaction rate (and therefore, cross section) was hypothesized, based on secondary reactions making DT neutrons.

There is circumstantial evidence that it was Ruhlig's 1938 paper that inspired Konopinski's DT insight, beyond the fact that he was a young diligent researcher who would have read the literature. Firstly, Bethe is acknowledged in a personal communication in this paper. Bethe was at Cornell, and at this time Konopinski  had a National Research Council fellowship under Bethe's supervision; they likely talked about this Michigan experiment. Also, Konopinski and Ruhlig probably knew each other: they were contemporaries at the University of Michigan in the early 1930s; Uhlenbeck was Konopinski's thesis advisor, and was also acknowledged in Ruhlig's thesis. 
Additional  evidence is in Oppenheimer's August 20, 1942, summary of the Berkeley meeting ``Memorandum on Nuclear Reactions''\cite{Chadwick:2022}, where he  describes the benefit of DT fusion. Surprisingly, he does not discuss adding tritium as another fuel to augment deuterium (it was thought too hard to manufacture tritium), and instead he described the energy production benefits of DD reactions breeding tritium on the fly, followed by {\it secondary} DT reactions, a suggestion credited to Bethe\cite{Hawkins:?}. That this is exactly the same kind of process that Ruhlig observed in the lab makes it likely that Konopinski had described Ruhlig's experiment to Oppenheimer and the other luminaries at Berkeley. It is unfortunate that neither Konopinski, nor Bethe, Teller or Oppenheimer, ever 
explicitly cited this pre-war evidence.

\begin{figure}[htbp]
\begin{center}
\includegraphics[width=3.1in]{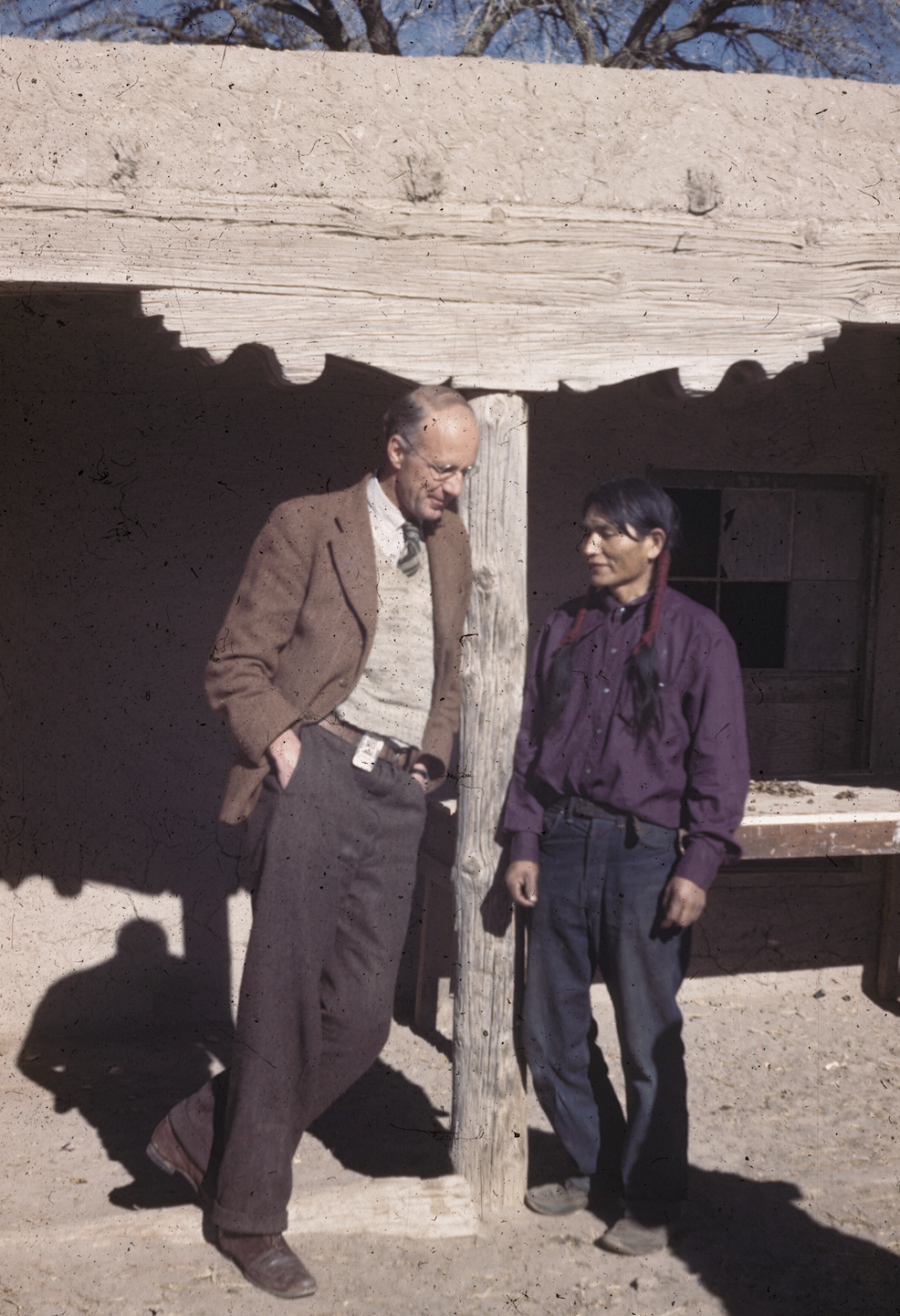}
\caption{Egon Bretscher at the San Ildefonso Pueblo, NM, January, 1946. Bretscher was the first to measure the DT cross section at relevant application energies, at Los Alamos 1945-6.
 (courtesy Churchill Archives Center, Cambridge University).}
\label{fig:bretscher-pic}
\end{center}
\vspace{-1.0cm}
\end{figure}

\section{DT Cross Sections in the 1940s -- 1950s}

The summer and fall of 1942 saw the creation of the Manhattan Project to design and build an atom bomb for the war effort. Although the focus was on fission, Oppenheimer's Berkeley meeting concluded 
with surprising optimism for the development of a thermonuclear bomb: 
``We have reached the conclusion that there is one method that 
 can be guaranteed to work'', though there were many challenges that they had not anticipated and it would take a decade to make the H-bomb a reality. Bethe recognized that a priority had to be the measurement of the DT cross section, and Purdue University was contracted by the Manhattan Project to take on this task, using their cyclotron. Baker and Holloway joined Schreiber and King at Purdue, and did the measurement in 1943.

A difficulty was producing enough tritium for the measurement. Oppenheimer and Bethe had instructed Berkeley's cyclotron team to produce the tritium using a deuterium beam incident on heavy water, via the D(d,p)T reaction. After a heroic effort, Segre, Kahn and  Kamen were able to send the tritiated heavy water to Purdue. Holloway sent a letter  to Berkeley complaining that they received far fewer tritons than were promised, but the reply came back that the sample they received was the best that could be made and that they should proceed. Measurements were made using accelerated tritons incident on deuterons in 
a heavy water (ice) target. This thick target experiment required an understanding of hydrogen-ion stopping powers to extract the cross
sections. Fortunately these had been measured in Germany in 1930 by Gerthsen\cite{Chadwick:2022}, and the world's expert on the theory of stopping powers was Bethe!

The Purdue experiment measured the $^3$He(d,p)$\alpha$ cross section (also for the first time ever), and then proceeded to measure  T(d,n)$\alpha$.
The cyclotron beam energy had to be tuned down to lower energies, but still the reactions were at higher energies than ideal for applications. A very large
cross section was found. Schreiber said: ``Dr. Hans Bethe was in frequent communication with us about these measurements and for a time he
simply did not believe our results''\cite{Chadwick:2022}: this was a surprise despite assessments by Rutherford and Ruhlig that the DT reaction could be \textit{exceedingly} probable. The $^3$HeD cross
section was also high, but Coulomb-suppressed relative to DT. The Manhattan Project Los Alamos 1943 reports documenting the
experiments\cite{baker:1943tr,baker:1943he} just factually present the results. There was no documentation  on a possible resonance enhancement;
this would come later, following subsequent experiments in 1945-6 at Los Alamos by Bretscher\cite{Bretscher:1946a}.  The Purdue University
data are shown in Fig.~\ref{fig:dt-dd-cf-efb80} as ``PU(1943)''; there are two sets of data points, denoted by circles, that correspond to
detectors at $90^\circ$ (higher-valued data) and $0^\circ$ (lower-valued data). These are compared to the current, most comprehensive
evaluation from the EDA R-matrix analysis (solid line) and, additionally, to the later (1946 and 1952) Los Alamos (LA) measurements. The PU
data proved to be quite good, considering that they are the first-ever DT measurement, although they did not extend down to the low energies
needed for fusion applications. They found a resonant peak cross section of 2.8 barns (versus the known value today of 5 barns) at a center-of-mass energy 
of about 130 keV (a factor of two too high compared to the value we know today).

Two years later, from 1945-6, the DT cross section was measured again, and more accurately and to lower energies, this time at Los Alamos. 
Teller had a group of researchers in Fermi's F Division focusing on the thermonuclear problem. 
As part of the British contingent of about twenty five scientists who joined the Los Alamos effort, Egon Bretscher  (Fig.~\ref{fig:bretscher-pic}) and his student Anthony French joined Fermi's division and were tasked to measure the DT and DD 
cross sections. Bretscher was already an established researcher at Cambridge, known for his accelerator measurements of the fission cross sections of uranium, and his use of source neutrons created in DD reactions. The tritium came from Oak Ridge's  Clinton Pile, although they only had ``a few cc'', and they used a heavy water (ice) target. Bretscher built a kenotron rectifier accelerator and was able to measure the cross sections at more relevant fusion energies, down to 6 keV c.m. energy. They  found a staggering result\cite{Bretscher:1946a,Bretscher:1946b} : the DT cross section was {\it one hundred times} larger than DD. Again, there was a surprise to see such a large DT resonant enhancement -- Hawkins'  1947 Manhattan District History recounted that ``the tritium cross section, however, was very much larger than had been anticipated at energies of interest''\cite{Hawkins:?} --
perhaps because the measurements extended so low in energy relative to both the DT Coulomb barrier which is about 300 keV\cite{Chadwick:2022} and to the resonance energy (the Purdue data might have suggested to them a  higher resonance energy than its true value, see Fig.~\ref{fig:dt-dd-cf-efb80}). Some years later, French remembered: 
``Then came the day when we, for the first time, put the tritium into the source of our accelerator in place of deuterium. And saw an absolute rain of pulses on the oscilloscope and we thought at first that something must be breaking down electrically because there were so many pulses. But we checked and it was for real ... So that was an amazing thing.''\cite{French:1992} 
His results are shown in Fig.~\ref{fig:dt-dd-cf-efb80} as ``(LA 1946)'' and again they compare  favorably to today's best R-matrix evaluations shown as solid lines, though are 
low for DT at the lowest energies. 

This time the Los Alamos results were accurate enough to infer a resonance effect and determine the DT resonance's location, which was estimated to be in the 49 -- 137 keV range in the 
c.m. energy frame, consistent with the known value today of 65 keV. Bretscher developed formulas for 
the DT cross section that combined Gamow's penetrability with a Breit-Wigner resonance 
description\cite{Bretscher:1949tdn}.

The very large DT cross section that peaks at 5 barns, with the resonance enhancement discovery, was a game changer, making thermonuclear fusion 
technologies possible. Dick Taschek \cite{Diven:1983} referred to it as ``the most important breakthrough'' on fusion made by the Manhattan Project. Bretscher's 1945-6 data\cite{Bretscher:1946a,Bretscher:1946b} were initially classified, but were subsequently released for open publication in the Physical Review in 1949\cite{Bretscher:1949tdn}.

In the early 1950s new capabilities were developed to measure the DT cross section yet more accurately, see the ``(LA 1952)'' data in Fig.~\ref{fig:dt-dd-cf-efb80}. This was the ``APSST'" experiment,
named after the Los Alamos authors Arnold, Phillips, Sawyer, Stovall and Tuck\cite{apsst:1952dtn,apsst:1954pr}, using a thin tritium gas target. Some of the innovations were suggested by Fermi and Garwin during their summer visit to Los Alamos. This experiment showed that the low-energy DT cross section was  higher still than 
the earlier Bretscher measurement (40\% higher for c.m. energies in the 6-10 keV range), a point that Teller enthusiastically emphasized\cite{Teller:1951}  in 1951.
Today, this experiment is viewed as the earliest high-accuracy DT measurement, for example in Bosch and Hale's reviews\cite{Bosch:1990,Bosch:1992bh}. 
Today's best R-matrix analysis of the DT A=5 nuclear reaction system, which is the basis for the latest ENDF evaluations, was influenced by these APSST data (Fig.~\ref{fig:dt-dd-cf-efb80}), although the later Los Alamos 1980s measurements
by Jarmie and Brown\cite{jarmie:1984prc} are most influential and still define our best understanding.

test

\section{Fermi's Astrophysical S-factor}

\begin{figure}[htbp]
\begin{center}
\includegraphics[width=3.in,origin=-0.5cm]{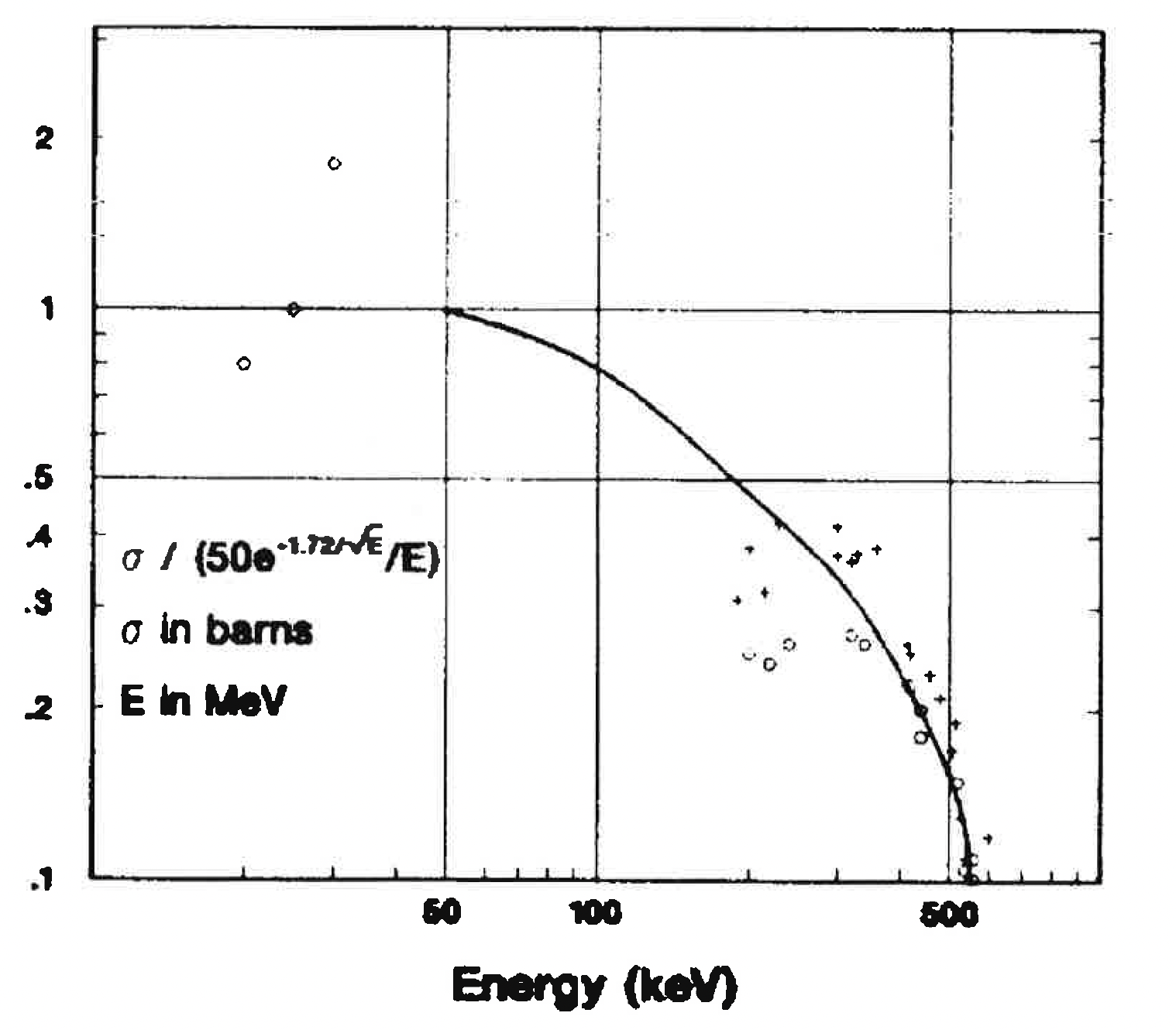}
\caption{A reproduction of Fermi's figure from his Los Alamos 1945 lectures, showing the DT cross section versus the laboratory triton energy. The DT cross section has the 1/E and exponential penetrability terms divided out, as is done in an S-factor representation. The three experimental points at the lower energies, near 20-30 keV, were 
preliminary (1945) Los Alamos Bretscher values; the higher data set around 200--600 keV are the 1943 Purdue Baker and Holloway values, 
also shown in Fig.~\ref{fig:dt-dd-cf-efb80} in the c.m. frame.}
\label{fig:fermi-dt}
\end{center}
\end{figure}

In the course of our investigations\cite{Chadwick:2022}, we have found a Fermi publication from Los Alamos that appears to be the first introduction of the ``S-factor'', a concept widely used in nuclear astrophysics in the representation of nucleosynthesis cross sections. The idea is to divide  the rapidly-varying energy-dependent Gamow penetrability and the $1/E_{\rm inc}$ factors from the cross section, leaving the S-factor nuclear resonant enhancement effect. Astrophysicists look back to the 1952 paper by 
Salpeter\cite{Salpeter:1952} and the Burbidge {\it et al.} B2FH review\cite{Burbidge:1957} from 1957 for the introduction of the S-factor. However,
Fermi made this same partition in his 1945 Los Alamos lectures for the DT cross section\cite{Chadwick:2022}, see 
Fig.~\ref{fig:fermi-dt}.  That Fermi was the first to introduce the S-factor 
(though he did not use this name) is not too surprising given the breadth of insights that be brought to physics.

{\bf Acknowledgements.} We thank M. Pearson and J. Katz for the early 1950s astrophysics S-factor references, and C. Carmer, M. Bernardin, A. Carr, N. Lewis, B. Archer, T. Kunkle, R. Capote, A. Hayes, S. Finnegan, S. Alhumaidi for useful discussions. This article is released as Los Alamos report LA-UR-22-32324 (2022).

\section{Appendix}

Given the historical importance of the 1938 Ruhlig experiment that reported secondary DT fusions following a D+D experiment, and the usefulness of such data to validate our simulation codes, we have initiated a modern measurement of this phenomenon at the 
Triangle Universities Nuclear Laboratory (TUNL) tandem Van de Graaff accelerator in Durham, North Carolina. TUNL's facility is one of the DOE Nuclear Physics Division's Center of Excellence, and most of its applied research programs are supported by the NNSA Stockpile Stewardship Academic Alliance. Los Alamos and Livermore staff routinely perform 
experiments at TUNL. Here we just report on the very early stages of these measurements.

The NIF facility has been used to study DD reactions creating secondary DT reactions, for example in the Symcap implosion experiments N1308013\cite{Cerjan:2018,Sayre:2019}. These experiments can yield ratios of TD to DDn of the order 10$^{-2}$, much  larger than our calculation of this ratio for
the Ruhlig experiment, though this reflects the very different environments with different stopping powers: one in a hot burning plasma; the other in 
room temperature heavy phosphoric acid. In those studies, Livermore's code calculations match their measured ratio well, to about 13 -- 32\% depending on the plasma stopping power model used. The TUNL experiment provides validation data in a complementary environment to that of NIF.

A deuteron beam can be made at TUNL with a minimum energy of 1.8 MeV, which can be reduced to 0.5 MeV with a degrader foil. The use of various deuteron-loaded targets are being investigated: heavy phosphoric acid (like Ruhlig); heavy water; and also deuterium gas. Likewise, different neutron detection approaches are being investigated: activation foils; neutron time-of-flight detection with an organic scintillator; and we also plan to implement  Ruhlig's approach of measuring upscattered protons from the high energy neutron elastic scattering.

\begin{figure}[htbp]
\begin{center}
\includegraphics[width=3.1in,origin=-0.5cm]{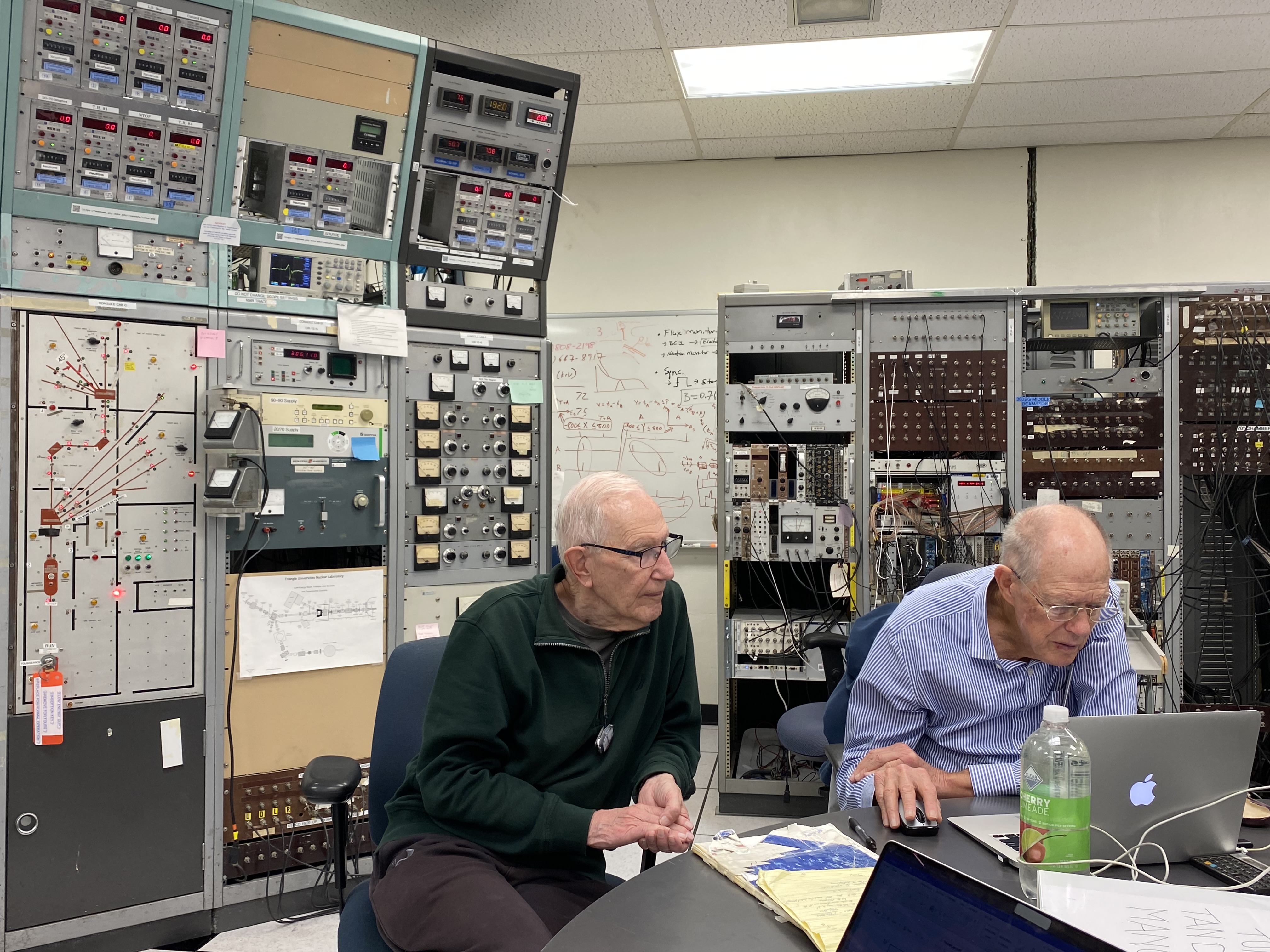}
\caption{Werner Tornow and Jerry Wilhelmy in the Van de Graaff control room at TUNL.}
\label{fig:tunl1}
\end{center}
\end{figure}

\begin{figure}[htbp]
\begin{center}
\includegraphics[width=3.1in,origin=-0.5cm]{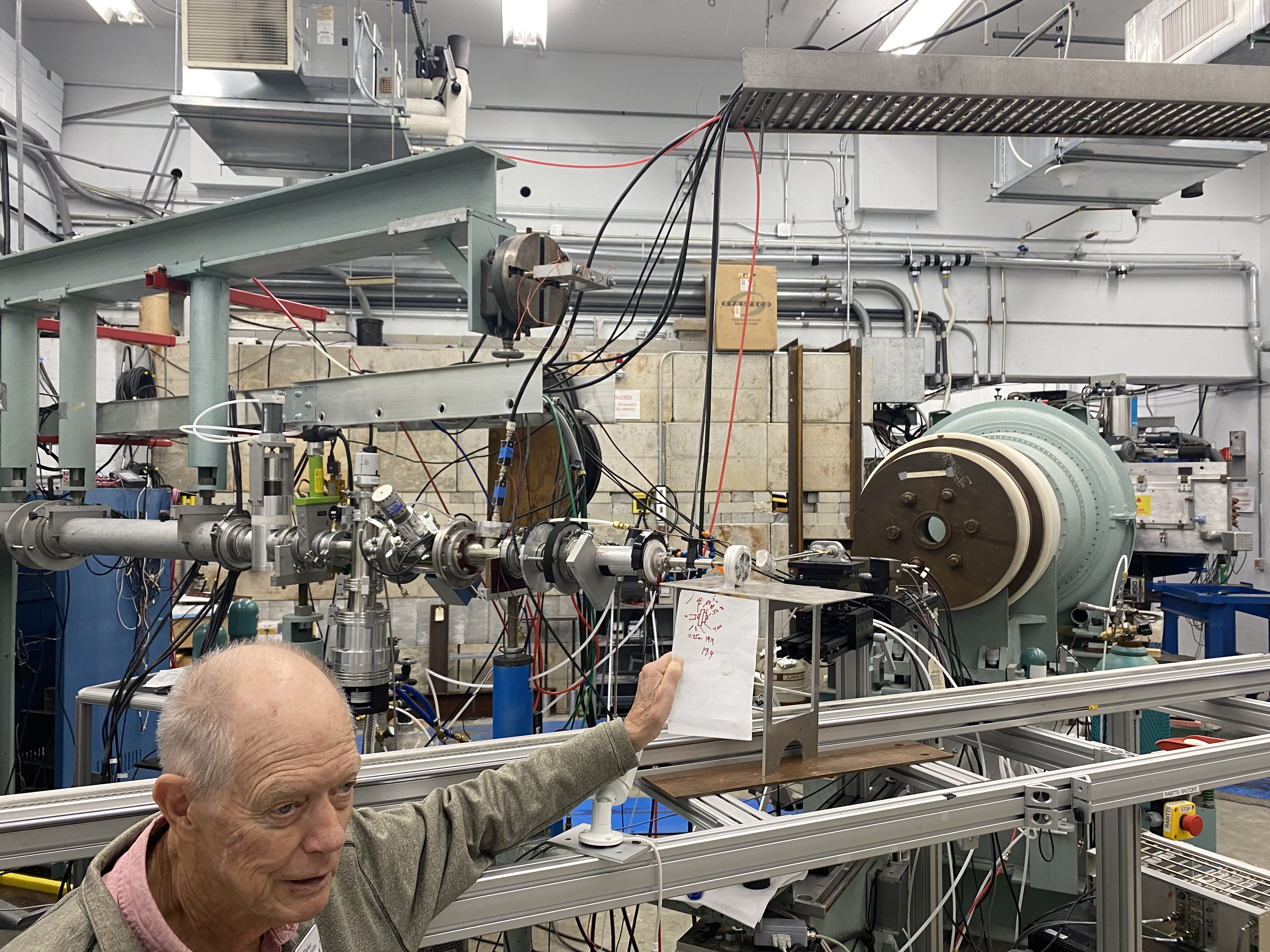}
\caption{The accelerator, showing the target cylinder container of D2 gas. Adjacent are the discs of activation ``foils'' to monitor the DDn and DTn neutrons; Jerry Wilhelmy is holding a diagram showing the foil stackup. The large spectrometer behind and  to the right of the target was not used in this run, but might be used in the future to provide neutron time-of-flight detection.}
\label{fig:tunl2}
\end{center}
\end{figure}

\begin{figure}[htbp]
\begin{center}
\includegraphics[width=3.2in,origin=-0.5cm]{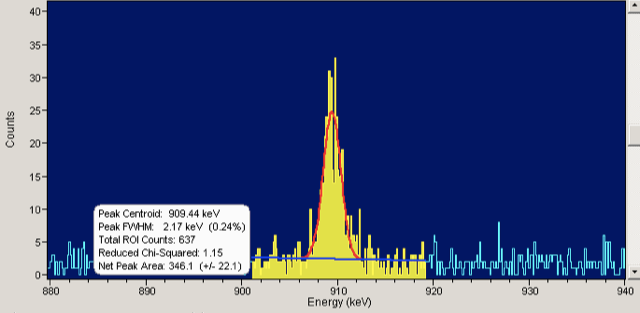}
\caption{Detection of secondary 14 MeV DT fusion neutrons following DD reactions at TUNL, using a Zr activation foil, mimicking Ruhlig's 1938 experiment. The decay gamma in $^{89}$Zr is shown, made through $\sim$14 MeV (n,2n) reactions. }
\label{fig:success}
\end{center}
\end{figure}

The goal of these experiments is to obtain quantitative data on DD reactions whose product T create secondary DT reactions, to compare against Ruhlig's observation and  to validate our understanding. For now, though, we just provide one first observation from the November 28, 2022 8-hour run using 1.75 $\mu$A 2.2 MeV deuterons (degraded to 1.4 MeV) on a gaseous deuterium target at four atmosphere pressure in a cylinder length 3 cm and diameter 1 cm. The deuteron projectile energy is higher than Ruhlig's 0.5 MeV because it was desired to initially create a larger number of primary DD fusions to improve measurement statistics; later, the energy will be reduced to 0.5 MeV, and the target will be changed to liquid heavy phosphoric acid and/or heavy water.
Activation foils\cite{Trkov:2020}: of Zr and Mg  measured the high energy 14 MeV DT neutrons  using threshold (n,2n) and (n,p) reactions, respectively, and an indium foil measured the more abundant DDn neutrons near 3 MeV using the (n,n')  reaction, see Figs.~\ref{fig:tunl1}, \ref{fig:tunl2}. A quantitative analysis of these results is ongoing; however, Fig.~\ref{fig:success} shows a 
$^{89}$Z decay gamma-ray decay signal in the Zr activation foil. Given the $^{90}$Zr(n,2n) 12.1 MeV threshold, this is a definitive observation of secondary DT fusion 14 MeV  neutrons.

\begin{figure}[htbp]
\begin{center}
\includegraphics[width=3.1in,origin=-0.5cm]{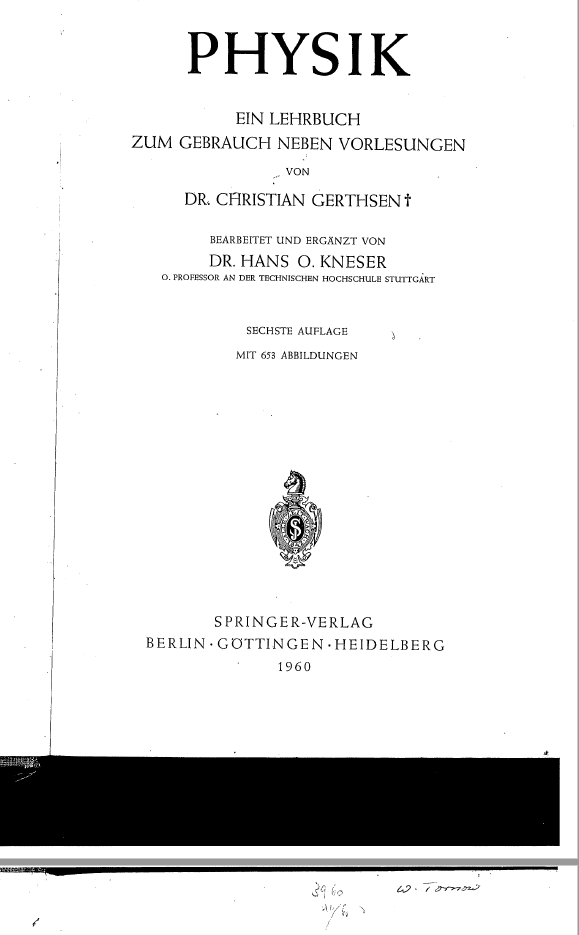}
\caption{Werner Tornow's Gerthsen textbook bought in 1961. in 1930 Gerthsen measured hydrogen stopping powers, which were used during the Manhattan project to first determine DT cross sections in thick-target experiments.}
\label{fig:Werner-textbook}
\end{center}
\end{figure}

One of us (Werner Tornow) comments: ``On a personal note: 57 years ago I used the DT reaction for the first time. Because our 2 MV Van de Graaff at Tuebingen was not pulsed, I was asked to implement the ``Associated Particle Method'', {\it i.e.}, use the alpha particle from the T(d,n)$\alpha$ reaction as a start signal for neutron time-of-flight measurements.'' Also, this paper had noted that hydrogen stopping power measurements dated back to Gerthsen in Germany, 1930. Fig.~\ref{fig:Werner-textbook} provides a screen shot of Tornow's copy of  Gerthsen's textbook, expensive at 39.60 German marks in 1961. Impressively, a 2015 edition is still in print!

\vspace{0.25in}
\noindent\rule{0.35\textwidth}{.4pt}


%


\newpage
\bibliographystyle{ans_js}   
\small\bibliography{bibliography.bib}  

\end{document}